\documentclass[a4paper,12pt]{article}

\newcommand{\RR}{{\rm R}}

\newcommand{\RRRR}{\RR^{1,3}}
\newcommand{\half}{{\textstyle{1\over2}}}

\newcommand{\polje}{\phi}
\newcommand{\polR}{\polje_{HV}}
\newcommand{\polA}{\polje_A}
\newcommand{\Polje}{\Phi}

\newcommand{\akcija}{I}
\newcommand{\akcijaR}{\akcija_{HV}}
\newcommand{\akcijaA}{\akcija_A}

\newcommand{\Lag}{{\cal L}}
\newcommand{\Lagf}{\Lag_{0}}
\newcommand{\Lagi}{V}

\newcommand{\LagRf}{\Lag_{HV0}}
\newcommand{\LagAf}{\Lag_{A0}}

\newcommand{\prop}{\widetilde{\Delta}}
\newcommand{\propF}{\prop_F}
\newcommand{\propA}{\prop_A}

\newcommand{\propR}{\prop_{HV}}
\newcommand{\unreg}{f_{HV}}
\newcommand{\altreg}{f_{A}}

\newcommand{\Gf}{G^{(n)}}
\newcommand{\GfZ}{\Gf_Z}
\newcommand{\GfR}{\Gf_{HV}}
\newcommand{\GfA}{\Gf_A}

\newcommand{\SmR}{S_{HV}}
\newcommand{\SmA}{S_A}

\title{Framework for finite alternative theories to a quantum field theory}
\author{Marijan Ribari\v c and Luka \v Su\v ster\v si\v c\thanks{Corresponding author. Phone +386 1 477 3258; fax +386 1 423 1569; electronic address: \tt luka.sustersic@ijs.si\rm} \\Jo\v zef Stefan Institute, p.p.3000, 1001 Ljubljana, Slovenia }
\date{}

\begin{document}

\maketitle

\begin{abstract}
The proposed framework is intended to faciliate the search for Lagrangian-based regularizations that introduce only such coefficients which are not regarded as formal auxiliary parameters that will eventually be got rid of during renormalization. Using the path-integral formalism, we put forward a new framework for such finite, alternative quantum theories to a given quantum field theory that have no formal auxiliary parameters. The corresponding alternative, perturbative Green functions are acceptably regularized perturbative expansions of the original Green functions, causal, and imply no unphysical free particles. To demonstrate that the proposed framework is feasible, we provide an alternative to the quantum field theory with $\polje^4$ interaction.
\end{abstract}

\vfill
Suggested short title: Framework for finite alternative theories ...
\vfill
PACS number: 11.90.+t \hfill UDC 5.30.145.7
\vfill
\vfill\eject

\section{Introduction}
\label{secintro}

In this paper, we will put forward a framework for constructing such alternative theories to a given quantum field theory (QFT) that provide solutions to the problem of ultraviolet divergencies \it without using formal, auxiliary parameters. \rm For simplicity, we will consider in some detail solely the case of the QFT of a single, self-interacting real scalar field $\polje(x)$ with $\polje^4$ interaction. In this case, the conventional, QFT, connected $n$-point Green functions scaled by a positive coefficient $Z^{-n/2}$, $\GfZ(x_1,\ldots, x_n)$, can be formally expressed as functional derivatives of a generating functional, i.e.,
\begin{equation}
  \GfZ(x_1,\ldots, x_n) = \left. { (-i)^{n} \delta^n \ln Z[J] \over\delta J(x_1) \cdots \delta J(x_n) }\, \right|_{J = 0} 
  \label{Greendef}
\end{equation}
with the generating functional
\begin{equation}
   Z[J] = \int \prod_{x} d\polje(x) \, e^{ i \akcija[\polje,J] } \,,
  \label{genfunc}
\end{equation}
where the action functional
\begin{equation}
   \akcija[\polje,J] = \int d^4x \Bigl ( \Lagf(\polje, \partial_\mu \polje) - \Lagi(\polje) + \polje J \Bigr) \,;
   \label{QFTaction}
\end{equation}
the QFT free-field Lagrangian 
\begin{equation}
   \Lagf = - \half (\partial_\mu \polje)^2 - \half m^2 \polje^2 \,;
   \label{QFTfreelagr}
\end{equation}
the interaction Lagrangian
\begin{equation}
   \Lagi(\polje) = {\lambda \over 4!} \polje^4 + {Z^2\lambda_0 -\lambda \over 4!} \polje^4 + {Z - 1\over 2} (\partial_\mu \polje)^2 
               + {Z m_0^2 - m^2\over 2} \polje^2 \,;
   \label{QFTintlagr}
\end{equation}
$m^2$, $m_0^2$, $\lambda$, and $\lambda_0$ are positive coefficients; the external source $J(x)$ is an arbitrary, real scalar field; we use the $(-1, 1, 1, 1)$ metric. An expression for the QFT scattering matrix of the quantum system, characterized by the action functional $\akcija[\polje,J]$, can be obtained in terms of $\GfZ$ by means of the Lehmann-Symanzik-Zimmermann (LSZ) reduction formula, see e.g. Sec.~10.3 in \cite{Weinberg}.

We can formally rewrite the generating functional (\ref{genfunc}) as
\begin{eqnarray}
   Z[J] &=& \exp \left \{ -i \int d^4x \, \Lagi( -i\delta/ \delta J ) \right \} \nonumber\\
        & & \qquad\times \int \prod_{x} d\polje(x) \, \exp \Bigl \{ i \int d^4x \left( \Lagf(\polje, \partial_\mu \polje) + \polje J \Bigr) \right\} \,.
  \label{genfuncexp}
\end{eqnarray}
Let $\polje_c(x)$ be the Feynman-Stueckelberg solution to the Euler-Lagrange equations of the action $\akcija[\polje,J]$ with $\Lagi = 0$. On changing the variable of functional integrals in (\ref{genfuncexp}) from $\polje$ to $\polje - \polje_c$, (\ref{genfuncexp}) can formally be written as
\begin{eqnarray}
   Z[J] &=& N \, \exp \left \{ -i \int d^4x \, \Lagi( -i\delta/ \delta J ) \right \} \nonumber \\
        & & \qquad\times  \exp \left \{ i \int d^4x d^4y \, \half J(x) \Delta_F(x-y) J(y) \right \}  \,,
   \label{genfunctrans}
\end{eqnarray}
where $N$ is a $J$-independent factor, irrelevant for the calculation of perturbative expansions of $\GfZ$, and the Feynman propagator $\Delta_F(x)$ is defined by the relation
\begin{equation}
   \polje_c(x) = \int  d^4y\, \Delta_F(x-y) J(y) \,,
   \label{propdefin}
\end{equation}
see e.g.~Sec.~1.2 and App.~B in \cite{Cheng}. By (\ref{propdefin}), (\ref{QFTaction}), and (\ref{QFTfreelagr}), the propagator $\Delta_F(x)$ is the Fourier transform of 
\begin{equation}
   \propF(k) = {1 \over k^2 + m^2 - i\epsilon} \qquad\hbox{with}\quad \epsilon \searrow 0 \,.
   \label{origprop}
\end{equation}

We can obtain the perturbative expansions of $\GfZ$ by replacing $Z[J]$ in (\ref{Greendef}) with (\ref{genfunctrans}), and then expanding in powers of the four coupling coefficients $\lambda$, $Z^2\lambda_0 -\lambda$, $Z - 1$, and $Z m_0^2 - m^2$. These expansions lead to the conventional Feynman rules for this theory \cite{Weinberg,Cheng}. In general, integrals over independent loops diverge, and perturbative expansions of $\GfZ$ are not well defined. Regularizing such divergent integrals in perturbative expansions of $\GfZ$, and then using a renormalization scheme and removing the cut-off and its auxiliary parameters, we can calculate the perturbative expansion of the QFT scattering matrix by the LSZ reduction formula. There is an infinite variety of \it acceptable regularizations \rm, i.e. such that the above procedure yields the same, renormalization-scheme-dependent, QFT result.

There is a widespread opinion that ultraviolet divergencies encountered in perturbative expansions of Green functions of the standard model are due to \it the inadequate description of ultra-high-energy phenomena by the present Lagrangian. \rm Which suggests that we should modify this Lagrangian to get rid of ultraviolet divergences. However, as noted e.g. in \cite{Weinberg}, in a QFT each complete, momentum-space, spin 0 Feynman propagator has the K\"all\'en-Lehman spectral representation with a non-negative spectral function, i.e., it is equal to
\begin{equation}
   \int_0^\infty {\rho(s)\over k^2 + s - i\epsilon}\, ds
   \label{Kahlenrep}
\end{equation}
with some $\rho(s) \ge 0$. That implies that in a QFT, also a complete spin 0 propagator cannot vanish faster than $\propF(k)$ as $| k^2 | \to \infty $. So it follows that in a QFT we can regularize none of the perturbative expansions of Green functions by modifying its Lagrangian. \it Some departure from the conventional framework of QFTs seems necessary to this end, but of course, it has to be compatible with the experimentally verified perturbative results of QFTs. \rm Which presents a sixty-years-old theoretical problem (see e.g. Sec.~1.3 in \cite{Weinberg}, and \cite{Villars}).

We believe that there is a Lagrangian-based theory of fundamental interactions such that we can regard perturbative expansions of its Green functions as acceptably regularized perturbative expansions of the Green functions of the standard model. However, up to now each candidate for such a theory that provides a non-perturbative regularization abandons some of the conventional properties of models in contemporary physics. Though upon removal of a cut-off during renormalization, these properties are restored in perturbative results. E.g., every version of the Pauli-Villars method proposed so far introduces auxiliary, unphysical particles with negative metric or wrong statistics; lattice regularization is based on discrete space-time; and string theories need to introduce additional space-time dimensions to avoid anomalies.

We intend to alleviate this conceptually displeasing situation where models have expected properties only when the cut-off is removed. To this end, we will generalize the 't Hooft-Veltman, Lagrangian-based version of the Pauli-Villars method (HV method), see Sec.~5 in \cite{Hooft}, to a continuous infinity of auxiliary fields but without introducing additional, \it auxiliary free particles. \rm We will \it argue \rm on the analogy of the above path-integral formalism (\ref{Greendef})--(\ref{origprop}). The key will be the LSZ reduction formula as, in general, \it it does not imply a direct connection between the number of free fields and the number of free particles. \rm 

For any given QFT, we will introduce a new, Lagrangian-based framework for constructing finite, alternative quantum theories to a QFT (FAQTs) such that: (i)~The action functional, $\akcijaA$, would equal the QFT action functional were it not for its free-field Lagrangian. (ii)~$\akcijaA$ \it is local in space-time and invariant under inhomogeneous Lorentz transformations. \rm (iii)~The FAQT connected $n$-point Green functions, $\GfA$, are defined by $\akcijaA$ through a generating functional on the analogy of (\ref{Greendef}). (iv)~The FAQT scattering matrix, $\SmA$, is defined in terms of $\GfA$ by the means of the LSZ reduction formula, and \it all of its coefficients are regarded as observable quantities. \rm (v)~Free particles of a FAQT, which are determined by the poles of $G^{(2)}_A$, are \it the same \rm as the free particles of the QFT: there are no additional, unphysical, auxiliary particles. (vi)~Perturbative expansions of $\GfA$ can be regarded as such acceptably regularized perturbative expansions of the QFT Green functions that are \it invariant under the same symmetry transformations as the QFT scattering matrix. \rm Hence, each FAQT provides at least as adequate a perturbative description as the QFT.

In Sec.~\ref{secunireg} we give a short description of the HV method for a single, self-interacting real scalar field in terms of the generating functional. In the HV method,  't Hooft and Veltman (i)~introduced a countable set of auxiliary fields, (ii)~altered only the QFT free-field Lagrangian, and (iii)~introduced Feynman-like rules that equal the conventional Feynman rules but for an altered propagator, which we label as the HV propagator. We could use the HV method to construct FAQTs were it not for its auxiliary, unphysical free particles.

In Sec.~\ref{sechighvar} we collate those properties of a HV propagator that are essential for an acceptable covariant regularization. We regard as \it an alternative \rm to $\propF$ each HV propagator that has these properties but no additional singularities which imply additional, free particles.

In Sec.~\ref{secproof} we give an example of how one may generalize the HV method to a continuous infinity of auxiliary fields so that the corresponding HV propagator is an alternative to $\propF$. So we construct a Lorentz invariant FAQT to the QFT of a single, self-interacting real scalar field. Thereby we show that the generalized HV method actually provides a feasible framework for constructing FAQTs to a QFT.

\section{Method of unitary regulators}
\label{secunireg}

't Hooft and Veltman invented, in addition to dimensional regularization, also the method of unitary regulators (HV method), which is a Lagrangian-based Pauli-Villars method with a discrete spectrum of auxiliary masses (see Secs.~2 and 5-8 in \cite{Hooft}). It seems that the HV method  has not been much noticed since we found no reference to it in Science Citation Index, though their report \cite{Hooft} itself has over 200 citations. Here we give an outline of a new, \it non-perturbative formulation \rm of the HV method in terms of a generating functional in the case of a single, self-interacting real scalar field with the QFT action functional $\akcija[\polje,J]$.

\it (a) The HV connected $n$-point Green functions. \rm Using a countable set of auxiliary real scalar fields $\polje_i(x)$, $i = 1, 2, \ldots$, we define the \it HV action functional: \rm
\begin{equation}
   \akcijaR[\polje_i,J] = \int d^4x \, \Bigl( \LagRf(\polje_i, \partial_\mu \polje_i) - \Lagi(\polR) + \polR J \Bigr) \,,
   \label{Raction}
\end{equation}
where the HV free-field Lagrangian
\begin{equation}
   \LagRf = - \half \sum_{i} c_i^{-1} \left ( (\partial_\mu \polje_i)^2 + \Lambda M_i^2 \polje_i^2 \right ) \quad\hbox{and}\quad
   \polR(x) = \sum_i \polje_i \,,
   \label{Rfreelagr}
\end{equation}
with $M_i$ and $c_i$ real coefficients such that 
\begin{equation}
   \Lambda M_1^2 = m^2 \,, \quad c_1 = 1 \,, \quad \sum_i c_i = 0 \,, \quad \sum_i M_i^2 c_i = 0 \,, \quad \sum_i | c_i | < \infty \,,
   \label{concpogoj}
\end{equation}
and $\Lambda$ is a positive cut-off parameter. We may regard $\akcijaR[\polje_i,J]$ as being obtained from $\akcija[\polje,J]$ by first replacing the field $\polje$ with the sum $\polR$ of auxiliary fields $\polje_i$, and then modifying $\akcija$ by replacing $\Lagf$ with a Lorentz-invariant alteration $\LagRf$. So in $\akcijaR$ \it all fields $\polje_i(x)$ have the same external source $J(x)$ \rm contrary to QFT formalism.

We define the HV connected $n$-point Green functions, $\GfR$, through functional derivatives (\ref{Greendef}) of the generating functional
\begin{equation}
   Z_{HV}[J] = \int \prod_{x,i} d\polje_i(x) \, e^{ i\akcijaR(\polje_i, J) } \,.
  \label{unigenfunc}
\end{equation}
And we define the HV scattering matrix, $\SmR$, in terms of $\GfR$ by means of the LSZ reduction formula (see e.g. Sec.~10.3 in \cite{Weinberg}, and Sec.~2.5 in \cite{Hooft}).

\it (b) Perturbative expansions. \rm We change the variables of functional integrals in (\ref{unigenfunc}) from $\polje_i$ to $\polje_i - \polje_{ic}$, where $\polje_{ic}(x)$ are solutions to the Euler-Lagrange equations of the action functional $\akcijaR$ with $\Lagi = 0$. We then formally obtain
\begin{eqnarray}
   Z_{HV}[J] &=& N_{HV} \, \exp \left \{ -i \int d^4x \, \Lagi( -i\delta/ \delta J ) \right \} \nonumber\\
        & & \qquad\times  \exp \left \{ i \int d^4x d^4y \, \half J(x) \Delta_{HV}(x-y) J(y) \right \}  \,,
   \label{unigenfunctrans}
\end{eqnarray}
where $N_{HV}$ is a $J$-independent factor, and the \it HV propagator \rm $\Delta_{HV}(x)$ is defined by the relation
\begin{equation}
   \sum_i \polje_{ic}(x) = \int d^4y\, \Delta_{HV}(x-y) J(y) \,.
   \label{regpropdef}
\end{equation}
We choose such solutions $\polje_{ic}(x)$ that $\Delta_{HV}(x)$ is the Fourier transform of
\begin{equation}
   \propR(k) = \unreg(k^2 - i\epsilon) \propF(k) \qquad\hbox{with}\quad \epsilon \searrow 0 \,,
   \label{propdiff}
\end{equation}
where \it the regularizing factor \rm
\begin{equation}
   \unreg(z) =  1 + (z + m^2 ) \sum_{i=2} {c_i \over z + \Lambda M_i^2} \,.
   \label{unireg}
\end{equation}
By (\ref{propdiff})--(\ref{unireg}), the HV propagator $\propR(k)$ is a \it linear combination \rm of the Feynman propagators of the HV free-field Lagrangian $\LagRf$; and $\propR = O(|k^2|^{-3})$ as $|k^2| \to \infty$, by (\ref{concpogoj}).

Expanding expression (\ref{unigenfunctrans}) in terms of the four coupling coefficients $\lambda$, $Z^2\lambda_0 -\lambda$, $Z - 1$, and $Z m_0^2 - m^2$ in $\Lagi(\polR)$, we obtain through (\ref{Greendef}) the perturbative expansions of $\GfR$; the resulting expressions lead to the Feynman-like rules for the perturbative calculation of $\GfR$. In particular, the perturbative expansions of $\GfR$ are calculated by the same sums over the Feynman diagrams as the perturbative expansions of $\GfZ$, but with the Feynman propagator $\propF(k)$ replaced with the HV propagator $\propR(k)$ whereas the vertices remain unchanged, by (\ref{genfunctrans}) and (\ref{unigenfunctrans}). 't Hooft and Veltman \cite{Hooft} obtained such regularized perturbative expansion of $\GfZ$ by applying the Feynman rules to the Lagrangian  $\LagRf - \Lagi$. Due to the form of the interaction and source terms, they could sum up various contributions to yield the above perturbative expansion of $\GfR$.

Now let us consider the limiting values of the perturbative expansions of the HV connected Green functions $\GfR$ when we remove the regularization by limiting the cut-off parameter $\Lambda \to \infty$. To calculate these limiting values, we: (i)~use the Bogoliubov-Parasiouk-Hepp-Zimmermann method; (ii)~assume that $Z$, $m_0$ and $\lambda_0$ are power series in $\lambda$, with coefficients that depend on $\Lambda$; and (iii)~determine these coefficients so that, order by order of $\lambda$, perturbative expansions of $\GfR$ with respect to $\lambda$ tend towards finite values as $\Lambda \to \infty$, see e.g. Sec.~2.2 in \cite{Cheng}. As, by (\ref{concpogoj}), the regularization factor $\unreg$ is such that
\begin{displaymath}
   \sup_{|z| < z_0} |\unreg^{(n)}(z) - \delta_{n0}| \to 0 \quad{\rm as} \quad \Lambda \to \infty \quad\hbox{for any}\quad z_0 > 0 , \quad n = 0,1,\ldots, 
\end{displaymath}
and
\begin{equation}
   \sup_{\Lambda > \Lambda_0} \sup_{z > 0} | z^n \unreg^{(n)} (z)| < \infty \qquad\hbox{for a} \quad\Lambda_0 > 0 ,\quad n = 0,1,\ldots, 
   \label{regucond0}
\end{equation}
the perturbative expansions of $\GfR$ tend towards the corresponding renormalized perturbative expansions of $\GfZ$ as $\Lambda \to \infty$. As a consequence, the perturbative expansions of $\GfR$ are an acceptable regularization of the perturbative expansions of $\GfZ$.

The perturbative expansions of $\GfR$ are causal, free of ultraviolet divergencies, and the corresponding perturbative expansion of $\SmR$ is unitary to all orders in the coupling coefficients; but it involves unphysical particles \cite{Hooft}. In particular, the zero-order term in the perturbative expansion of $G_{HV}^{(2)}$ equals the HV propagator $\propR(k)$. So the poles of $\propR(k)$ determine the properties of free particles of the HV scattering matrix $\SmR$: (i)~the position of the pole determines the mass of the particle, and (ii)~when the residue is negative, the particle is considered unphysical (a ghost), since then transition probabilities can be negative in the tree order \cite{Hooft}. So, the HV scattering matrix $\SmR$ predicts free scalar particles with masses $M_i$ that are unphysical when $c_i < 0$. By (\ref{concpogoj}), the above regularization of $\GfZ$ by the HV method introduces additional free particles, at least one of which is unphysical.

\section{Alternatives to the QFT propagator}
\label{sechighvar}

We do not need the additional poles of the HV propagator $\propR(k)$ by themselves for an acceptable regularization of perturbative expansions of $\GfZ$. As some of them correspond to unphysical free particles we would prefer a HV method with such a HV propagator that has properties sufficient for an acceptable regularization of perturbative expansions of $\GfZ$ but no additional singularities. We will refer to such a HV propagator as an \it alternative \rm to the QFT spin $0$ propagator $\propF(k)$ and denote it by $\propA(k)$ provided
\begin{equation}
   \propA(k) = \altreg(k^2 - i\epsilon) \propF(k) \qquad \hbox{with}\quad \epsilon \searrow 0 \,,
   \label{alternative}
\end{equation}
where: (i)~the regularizing factor $\altreg(z)$ does not depend on $\epsilon$, and is an analytic function of complex variable $z$ except somewhere along the segment $z \le z_d < -m^2$ of the negative real axis. (ii)~$\altreg(-m^2) = 1$. (iii)~$\altreg(z)$ is real on the positive real axis; so $\altreg^*(z) = \altreg(z^*)$. (iv)~$\sup_z (1 + |z|^{3/2}) | \altreg(z)| < \infty$. So all integrals over independent loops converge, and also the Wick rotation is possible. (v)~The coefficients of $\altreg(z)$ can be made to depend on some cut-off parameter $\Lambda$ so that for any $\Lambda \ge \Lambda_0 > 0$ the regularizing factor $\altreg$ has properties (i)--(iv) and satisfies relations (\ref{regucond0}).

By (i)-(v), we can regard $\propF$ as a covariant, low-energy approximation to its alternative $\propA$; the coefficients of $\propA$ can be chosen to make this low-energy approximation as accurate as desired.

Using Cauchy's integral formula we can conclude that each $\propA(k)$ admits the K\"all\'en-Lehman representation (\ref{Kahlenrep}) with the spectral function
\begin{equation}
   \rho(s) = \delta(s-m^2) - \pi^{-1} (m^2 -s)^{-1} \lim_{y \to 0} \Im \altreg(-s + iy) \,,
   \label{KLrho}
\end{equation}
$s, y > 0$. By (i)-(iv), $\rho(s)$ is real, $\rho(s) = O(s^{-5/2})$ as $s \to \infty$, and
\begin{equation}
   \int_0^\infty s^\ell \rho(s)\, ds = 0 \qquad\hbox{for}\quad \ell = 0, 1;
   \label{KLcon}
\end{equation}
so $\rho(s)$ is changing sign, in contrast with the spectral function in QFT. By (\ref{Kahlenrep}) and (\ref{KLrho}), we may regard $\propA$ as a sum of $\propF$ and of a Pauli-Villars regulator with a continuous spectrum of auxiliary masses \cite{Pauli}.

Following 't Hooft and Veltman \cite{Hooft}, we can show that by replacing the QFT propagator $\propF$ in the perturbative expansions of $\GfZ$ with its alternative $\propA$ we obtain acceptably regularized, covariant, and causal expressions that equal the perturbative expansions of $\GfA$. Which result, by means of the LSZ reduction formula, in a covariant and unitary perturbative scattering matrix $\SmA$ that is without additional free particles but may involve additional, unphysical particles generated by the interaction \cite{Hooft}. However, as there are no additional particles at $\Lagi = 0$, there also are no additional particles in a vicinity of $\Lagi = 0$ if the zeroes of $(G_A^{(2)})^{-1}$ are continuous functions of the coupling coefficients in this vicinity. However, it is an open question when this is so.

In the next section we will generalize the HV method to auxiliary fields having a continuous, four-vector index so that the sum in the definition (\ref{unireg}) of the regularizing factor $\unreg(k^2)$ is replaced with an integral that has no poles as a function of $k^2$. Thereby we will avoid the additional free particles implied through the LSZ reduction formula by the additional poles of $\propR(k)$. So we will be able to demonstrate the existence of such a local and invariant action functional that the corresponding (in the sense of (\ref{genfunc})--(\ref{genfunctrans})) propagator is an alternative to $\propF$. Each such action functional defines a FAQT, as specified in Sec.~\ref{secintro}, to the QFT of a single, self-interacting real scalar field.

\section{An example}
\label{secproof}

One may wonder whether it is possible to generalize the HV method so that the HV propagator $\propR$ is an alternative to $\propF$, and thus construct a FAQT to the QFT of a single, self-interacting real scalar field with action functional (\ref{QFTaction}). The following action functional (\ref{Aakcija}) demonstrates that this is possible and so provides an \it example \rm of such a FAQT as specified in Secs.~\ref{secintro} and \ref{sechighvar}.

We introduce infinitely many auxiliary real scalar fields with a continuous, real four-vector index $p$, $\Polje(x,p)$; their weighted sum
\begin{equation}
   \polA(x) = \int d^4p\, f(p^2) \Polje(x,p) \,;
   \label{tranpolje}
\end{equation}
and the FAQT free-field Lagrangian 
\begin{equation}
   \LagAf =  \half \int d^4p\,\Polje(x,-p)[ p^\mu \partial_\mu + t(p^2) ] \Polje(x, p) + q \polA^2 \,,
   \label{AfreeLagr}
\end{equation}
where $q$ is a real coefficient; $f(y)$ and $t(y)$ are real functions of real $y$, yet to be specified, such that $f^2(y)/t(y)$ is finite at $y = 0$; and the integral
\begin{equation}
   \int d^4p \dots \;\; = -i \lim_{y\to \infty} \int_{-iy}^{iy} dp^0 \int _{p^2 \le y^2} d^3{\bf  p} \dots \;\; \,.
   \label{traninte}
\end{equation}
A possible physical significance of such a FAQT free-field Lagrangian is considered in \cite{mi001}.

We construct the FAQT action functional $\akcijaA$ by replacing $\Lagf$ with $\LagAf$, and $\polje$ with $\polA$ in the QFT action functional $\akcija$; i.e.,
\begin{equation}
   \akcijaA[\Polje, J] = \int d^4x \, \left [ \LagAf(\Polje, \partial_\mu\Polje) - \Lagi(\polA) + \polA J \right ] \,.
  \label{Aakcija}
\end{equation}
Note that we do not introduce an external source for each individual, auxiliary real scalar field $\Polje(x,p)$, but only for the combination $\polA$ that appears in the interaction Lagrangian $-V$. The FAQT action functional $\akcijaA$ is \it real, quadratic, local in spacetime, and invariant under the inhomogeneous Lorentz transformations \rm
\begin{equation}
   x \to L x + a \,, \quad   
   p \to L p \,, \quad   
   \Polje(x,p) \to \Polje(L x + a, L p) \,, \quad
   J(x) \to J(L x + a) \,.
   \label{lorentztrans}
\end{equation}

We define the FAQT connected $n$-point Green functions, $\GfA$, by functional derivatives (\ref{Greendef}) of the generating functional
\begin{equation}
   Z_A[J] = \int \prod_{x,p} d\Polje(x,p) e^{ i \akcijaA [\Polje, J] } \,.
  \label{Agenfunc}
\end{equation}
And we define the FAQT scattering matrix, $\SmA$, in terms of $\GfA$ by means of the LSZ reduction formula.

We change the variables of functional integrals in (\ref{Agenfunc}) from $\Polje(x,p)$ to $\Polje(x,p) - \Polje_c(x,p)$, where $\Polje_c(x,p)$ is a solution to the Euler-Lagrange equations of the FAQT action $I_A$ with $\Lagi = 0$. So we can  formally write the generating functional (\ref{Agenfunc}) as
\begin{eqnarray}
   Z_A[J] &=& N_A \, \exp \left \{ -i \int d^4x \, \Lagi( -i \delta/ \delta J ) \right \} \nonumber\\
        & & \qquad\times  \exp \left \{ i \int d^4x d^4y \, \half J(x) \Delta(x-y) J(y) \right \}  \,,
   \label{Agenfunctrans}
\end{eqnarray}
where $N_A$ is a $J$-independent factor, and the propagator $\Delta(x)$ is defined by the relation
\begin{equation}
   \int d^4p f(p^2) \Polje_c(x,p) = \int d^4y \, \Delta(x-y) J(y) \,.
   \label{tranpropdef}
\end{equation}
We can choose \cite{mi002} such a solution $\Polje_c(x,p)$ that the Fourier transform of $\Delta(x)$ is
\begin{equation}
   \prop(k) = { - I(k^2 - i\epsilon) \over 2q I(k^2 - i\epsilon) + 1 } \qquad \hbox{with}\quad \epsilon \searrow 0  \,,
   \label{tranprop}
\end{equation}
where $I(z)$ is the analytic function of the complex variable $z$ equal to
\begin{equation}
   2\pi^2 \int_0^\infty dy\, f^2(y) [ y/ t(y) ] \left( \sqrt{1 + z y/ t^2(y)} + 1 \right) ^{-1} \qquad\hbox{for}\quad z > 0 \,.
   \label{imnr}
\end{equation}

We can choose such $q$, $f(y)$, and $t(y)$ that the propagator $\prop(k)$ becomes an alternative to $\propF$; and so the action functional $I_A$ defines such a FAQT as specified in Secs.~\ref{secintro} and \ref{sechighvar}. To show by a simple example that this is feasible: (i)~Let 
\begin{equation}
   t(y) = (-1)^j v_j \sqrt{y} \qquad\hbox{for}\qquad y \in [j-1,j) \,, \quad j= 1, 2, 3, 4, 
   \label{examplet}
\end{equation}
where $v_j = [ 1 + (2j - 5)\eta)] \Lambda $, $\Lambda \ge \Lambda_0 > m/(1 - 3\eta)$, and $ \eta \in (0, 1/3)$. (ii)~Let 
\begin{equation}
   q = { - m^2 r_1 \over 2r_1 + m^2 r_{-1} } \,, \qquad \hbox{with}\quad r_n = \sum_{j=1}^4 d_j ( v_j^2 - m^2 )^{n/2} 
   \label{exampleq}
\end{equation}  
and $d_j = (-1)^j (4 - |2j - 5|)$. (iii)~Let $f(y)$ be such that 
\begin{equation}
   2\pi^2 \int_{j-1}^j f^2(y) \sqrt{y} \, dy = { |d_j | m^2 \over 2 q r_1 } \,, \quad j= 1, 2, 3, 4,
   \label{examplef}
\end{equation}
and let $f(y) = 0$ for $y > 4$. In such a case, we can check by inspection that: (i)
\begin{equation}
   \prop(k) = \altreg(k^2 - i\epsilon ) \propF(k)  \,, 
   \label{alternative2}
\end{equation}
with the regularization factor
\begin{eqnarray}
   \altreg(z) & = &   \Big [ \sum_{j=1}^4 d_j ( \sqrt{v_j^2 + z} + v_j )^{-1} \Big ] \Big/ \Big [ 2q \sum_{j=1}^4 d_j ( \sqrt{v_j^2 - m^2} + v_j )^{-1}
         \nonumber \\
                 & &\qquad \times ( \sqrt{v_j^2 + z} + v_j )^{-1} ( \sqrt{v_j^2 + z} + \sqrt{v_j^2 - m^2}\,\, )^{-1} \Big ] \,.
   \label{tranpropexample}
\end{eqnarray}
(ii)~$\altreg(z)$ is an analytic function of $z$ except along the segment $z \le z_d$, $z_d = - \Lambda^2 (1 -3\eta)^2 < -m^2 $, of the negative real axis for each $\eta \in (1/3 - 0.047, 1/3)$; to infer this, we take into account that $\Re \sqrt z \ge 0$ and that $\Lambda/(\sqrt{v_j^2 - m^2} + v_j) \to  \infty$ as $\eta \nearrow 1/3$ only if $j = 1$. 

(iii)~$\altreg(z)$ is real for $z \ge - v_1^2$, $\altreg(-m^2) = 1$, $\sup_z (1 + |z|^{3/2}) |\altreg(z)| < \infty$, and relations (\ref{regucond0}) are satisfied. Thus, $\prop(k)$ is an alternative to $\propF(k)$, with $\Lambda \ge \Lambda_0 $ as a cut-off parameter. 

Note that there are infinitely many functions $f(y)$ and $t(y)$ that result in the same alternative, (\ref{alternative2})--(\ref{tranpropexample}), to the propagator $\propF(k)$.

On the analogy with alternatives to the spin 0 Feynman propagator, we have defined also alternatives to the spin $\half$ Feynman propagator and to the Feynman propagator in unitary gauge for massive spin 1 bosons, and provided an example of the corresponding FAQT free-field Lagrangians \cite{mi003}: We used them to construct an example of a FAQT to QED with massive photons in the unitary gauge.

Similarly we can demonstrate that there are FAQTs to the standard model. However, it is an open question what properties the free-field Lagrangian of a FAQT to the standard model must have to make this theory: (i)~without additional particles created by the interaction; (ii)~of interest for the study of non-perturbative phenomena, cf. \cite{Gupta}, and (iii)~a better way of extracting physical data from the interaction Lagrangian of the standard model. In this connection it is not of primary importance how convenient are the resulting, acceptably regularized Green functions for further calculations---the main concern of all regularizations so far, cf. \cite{Deminov}.

\section{Some comparisons between the FAQT and QFT formalisms}
\label{seccomp}

From the preceding examples we note the following differences and similarities between the FAQT and QFT formalisms:

(a)~K\"all\'en-Lehman representation (\ref{Kahlenrep}) with the spectral function (\ref{KLrho}) of the FAQT propagator $\propA$ defined by (\ref{tranprop}) is a consequence of assumptions about the FAQT action functional $I_A$. These classical  assumptions do not imply the QFT premises on which the K\"all\'en-Lehman representation (\ref{Kahlenrep}) with a non-negative spectral function is based in QFTs, and do not contradict the positivity postulates of quantum mechanics, cf. Sec.~10.7 in \cite{Weinberg}.

(b)~According to Buchholz and Haag \cite{Buchholz} one can take as a postulate that each QFT ``is completely described by a finite number of covariant fields (each having a finite number of components)''. However, in constructing a FAQT we replace  each QFT field with a field that has a continuous infinity of components---an essential difference characteristic also of string theories.

(c)~If we replace the FAQT free-field Lagrangian $\LagAf$ in the FAQT action functional $\akcijaA$ with the QFT  free-field Lagrangian $\Lagf$ with $\polje=\polA$, the resulting action functional becomes equivalent to the QFT action functional $\akcija$ as far as the connected Green functions are concerned. So we may take that QFT of a single, self-interacting real scalar field and the FAQT based on $\akcijaA$ actually differ only in their free-field Lagrangian densities. The interaction and source parts of the FAQT action functional $\akcijaA$ are completely determined by the corresponding parts of the QFT action functional $\akcija$.

(d)~Contrary to the QFT free-field Lagrangian, \it a FAQT free-field Lagrangian by itself does not specify the free particles of a FAQT; \rm and the number of FAQT free particles is finite whereas the number of FAQT free fields is not.

\section{Summary}
\label{secsummary}

In this paper: (i)~we have put forward a framework for constructing such \it finite alternative quantum theories \rm (FAQTs) to a QFT that respect \it the conventional premises of theoretical physics, \rm and (ii)~we have given an example to prove that this framework is feasible.

The basic premises of the proposed mathematical framework for constructing a FAQT to a given QFT are that:
\begin{itemize}
\item[(i)]A FAQT can be characterized by a local, invariant action functional.
\item[(ii)]The FAQT, connected $n$-point Green functions are defined by functional derivatives of a generating functional. They are invariant under the same symmetry transformations as the QFT scattering matrix.
\item[(iii)]The FAQT scattering matrix is defined in terms of the FAQT Green functions by means of the LSZ reduction formula. Its coefficients are regarded as observable quantities: it has \it no auxiliary parameters. \rm
\item[(iv)]The FAQT free particles are the same as the QFT ones.
\item[(v)]Perturbative expansions of the FAQT Green functions may be regarded as such acceptably, covariantly regularized perturbative expansions of the QFT Green functions that do not depend on auxiliary parameters.
\end{itemize}
So each FAQT contains all perturbative results of the QFT in question. In particular, in the asymptote of a negligible cut-off the perturbative expansion of the FAQT scattering matrix equals that of the QFT scattering matrix.

We construct a FAQT to a QFT as follows:
\begin{itemize}
\item[(i)]We take the QFT Lagrangian and introduce external sources to form the action functional we need to specify its QFT connected Green functions through the generating functional.
\item[(ii)]In the QFT action functional, we replace each field but the external sources with an average over a continuous infinity of auxiliary fields of the same kind.
\item[(iii)]We replace the QFT free-field Lagrangian with such a local free-field Lagrangian that the resulting action functional becomes a FAQT action functional as specified above.
\end{itemize}
In such a case, the Feynman-like rules for calculating the FAQT Green functions are similar to the ones of the QFT in question: only the QFT propagators are replaced with FAQT propagators, which are determined through the FAQT action functional without the interaction part.

By providing an example in terms of functions defined in the eight-dimensional space $\RRRR \times \RRRR$, we have shown in Sec.~\ref{secproof} that the above framework, based on the path-integral formalism, is feasible.

\section*{Acknowledgement}

We are grateful to Matja\v z Polj\v sak for many helpful remarks and discussions.

\vfill\eject


\begin{thebibliography}{99}

\bibitem{Weinberg}S. Weinberg, \it The Quantum Theory of Fields, \rm Cambridge University Press, Cambridge (1995), Vol. I.

\bibitem{Cheng}T.-P. Cheng, L.-F. Li, \it Gauge Theory of Elementary Particle Physics, \rm Claredon Press, Oxford (1992).

\bibitem{Villars}F. Villars, in M. Fierz and V. F. Weisskopf (Eds.), \it Theoretical Physics in the Twentieth Century, \rm Interscience Publishers, New York (1960), p.~78; C. J. Isham, A. Salam, J. Strathdee, Phys. Rev. D \bf 5 \rm (1972) 2548;  T. Y. Cao, S. S. Schweber, Synthese \bf 97 \rm (1993) 33; L. M. Brown (Ed.), \it Renormalization, \rm Springer-Verlag, New York (1993).

\bibitem{Hooft}G. 't Hooft, M. Veltman, \it Diagrammar, \rm CERN report 73-9 (1973); reprinted in G. 't Hooft, \it Under the Spell of  Gauge Principle, \rm World Scientific, Singapore (1994).

\bibitem{Pauli}W. Pauli, F. Villars, Rev. Mod. Phys. \bf 21 \rm (1949) 434.

\bibitem{mi001}M. Ribari\v c, L. \v Su\v ster\v si\v c, hep-th/9810138; Fizika (Zagreb) B \bf 8 \rm (1999) 441.

\bibitem{mi002}M. Ribari\v c, L. \v Su\v ster\v si\v c, Fund. Phys. Lett. \bf 7 \rm (1994) 531.

\bibitem{mi003}M. Ribari\v c, L. \v Su\v ster\v si\v c, hep-th/0010209; Fizika (Zagreb) B\bf 10 \rm (2001) 73.

\bibitem{Gupta}M. Asorey, F. Falceto, Phys. Rev. D \bf 54 \rm (1996) 5290.

\bibitem{Deminov}M. M. Deminov, A. A. Slavnov, Phys. Lett. B \bf 501 \rm (2001) 297.

\bibitem{Veltman}M. Veltman, \it Diagrammatica, \rm Cambridge University Press, Cambridge (1994).

\bibitem{Buchholz}D. Buchholz, R. Haag, J. Math. Phys. \bf 41 \rm (2000) 3674.

\end{thebibliography}
\end{document}